\documentclass[twocolumn]{pasj00}

\begin{document}
\SetRunningHead{M. Takizawa}
               {X-Ray and Mass Distribution in the Merging Galaxy
               Cluster 1E 0657-56}
\Received{}
\Accepted{}

\title{On the X-Ray and Mass Distribution in the Merging Galaxy 
       Cluster 1E 0657-56:
      Ram Pressure-Stripping in Substructures with an NFW Density Profile}

\author{Motokazu \textsc{Takizawa}}
\affil{Department of Physics, Yamagata University, 
Kojirakawa-machi 1-4-12, Yamagata 990-8560}
\email{takizawa@sci.kj.yamagata-u.ac.jp}

%


\maketitle

\begin{abstract}
 We investigate the X-ray and mass distribution in the merging galaxy
 cluster 1E 0657-56.
 We study head-on collisions of two virialized clusters with an NFW
 density profile in the $\Lambda$CDM universe using an $N$-body +
 hydrodynamical code. 
 A clear off-set of an X-ray peak from a mass peak, which is like what is 
 reported in \citet{Clow04}, is first reproduced in the $N$-body +
 hydrodynamical simulations.
 We estimate the ram pressure-stripping conditions of the
 substructure in mergers of two NFW dark halos using a simple analytical 
 model. We find that the ram pressure dominates the gravity of
 the substructure when the smaller cluster's mass is less than 
 approximately one tenth of the larger cluster's mass.
 The characteristic X-ray and mass structures found in 1E 0657-56 suggest
 that neither the ram pressure nor the gravitational bound force
 overwhelms the other and that the mass ratio between the progenitors 
 is near the critical value mentioned above.
\end{abstract}

\section{Introduction}
According to the standard scenario of cosmological structure formation, 
clusters of galaxies form through mergers of smaller subclusters.
In fact, some of them are forming now. Merging clusters are the sites 
of structure formation in the universe that can be investigated in detail
via different types of observations.
Mergers play important roles in cluster evolution itself.
Cluster mergers cause bulk flow motion, turbulence, 
shocks, and/or contact discontinuities in the intracluster medium (ICM), 
which give us clues 
to investigate various physical processes in the ICM 
(see \cite{Sara02} for a review). 
Turbulence and shocks most likely play crucial roles in particle
acceleration in the ICM (e.g. \cite{TaNa00}; \cite{Ohno02}; 
\cite{Taki02}; \cite{Fuji03}; \cite{Buru03} ). 
Strong bulk flow motion and violent pressure changes in the ICM during
mergers may affect
star formation activities of the member galaxies (\cite{Fuji99};
\cite{Bekk03}).
$N$-body + hydrodynamical numerical simulations have been carried out 
to study cluster merger physics 
(e.g. \cite{Schi93}; \cite{Ishi96}; \cite{Roet97}; \cite{Taki99}, 
\yearcite{Taki00}; \cite{Rick01}; \cite{Ritc02}).
Comparison of such numerical simulations
with different kinds of observations give us deep insight into 
the cluster physics.

1E 0657-56 is one of the most well-known examples of a merging cluster. 
It is the hottest known cluster and has a very powerful radio
halos \citep{Lian00}. 
Its simple geometry makes this cluster one of the most suitable case
to investigate cluster merger physics. There are two peaks in both the X-ray
surface brightness distribution \citep{Mark02} 
and galaxy distribution \citep{Barr02}, but
their positions do not agree with each other. 
Observations of the line-of-sight 
velocities of the member galaxies suggest
that its collision axis is almost perpendicular to the line-of-sight
\citep{Barr02}.
From X-ray observations, a bow shock and a cold front are 
found in front of the smaller subcluster,
which suggest that its Mach number is $2 - 3$ \citep{Mark02}.

Recently,  \citet{Clow04} investigated the mass 
distribution in 1E 0657-56 through 
weak gravitational lensing. They show clear offsets of the mass
density peaks from the X-ray peaks, and that the mass distribution is quite 
similar to the galaxy one. The smaller substructure in mass is 
ahead of the X-ray one. They claim that this structure occurs because
ICM experience ram pressure but dark matter (DM) and galaxies do not. 
Although the above-mentioned naive ram pressure-stripping scenario seems
to be correct, such characteristic off-sets of X-ray peaks to mass peaks
have never been reported in the past numerical simulations. 
In this paper, we show the first results that successfully
reproduce such characteristic structures in the $N$-body + hydrodynamical 
simulations, and discuss their implications using a simple analytical model.

The rest of this paper is organized as follows. In \S 2 we describe 
the adopted numerical method and initial conditions for our simulations.
In \S 3 we present the simulation results. In \S 4 we show simple analytical
estimation for ram pressure-stripping conditions in mergers of two
clusters with an NFW density profile. In \S 5 we summarize the results.

\section{The Simulations}

\subsection{The Numerical Method}
In the present study, we consider clusters of galaxies consisting of 
two components: collisionless particles corresponding to the galaxies and 
DM, and gas corresponding to the ICM. When calculating gravity, 
both components are considered, although the former dominates over 
the latter. Radiative cooling and heat conduction are not included. 
We use the Roe total variation diminishing (TVD) scheme to solve the
hydrodynamical equations for the ICM 
(see \cite{Hirs90}). The hydrodynamical part of the code used here 
is identical with what is used in \citet{Taki05}. 
Gravitational forces are calculated by the Particle-Mesh (PM) method 
with the standard FFT technique for the isolated boundary conditions
(see \cite{Hock88}).
The size of the simulation box and the number of the grid points are 
11.8 Mpc $\times$ (5.92 Mpc)$^2$ and 256 $\times$ (126)$^2$,
respectively. The total number of the $N$-body particles used in the
simulations is 256 $\times$ (128)$^2$,
which is approximately $4.2 \times 10^6$.

\subsection{Models and Initial Conditions}
We consider mergers of two virialized subclusters with an NFW
density profile \citep{Nava97} in the $\Lambda$CDM
universe ($\Omega_0=0.25$, $\lambda_0=0.75$) for DM.
DM masses of the larger and smaller subclusters are 
$1.00 \times 10^{15} M_{\odot}$ and $6.25 \times 10^{13} M_{\odot}$,
respectively. Thus, the mass ratio is 16:1. 
The initial density profiles of the ICM are assumed to be those of a
beta-model, where the core radius is a half of the scale radius of the DM
distribution, and $\beta=0.6$. The gas mass fraction is set to be 0.1
inside the virial radius of each subcluster.
The parameters such as a virial radius $r_{\rm v}$, 
concentration parameter $c$
for each subcluster are summarized in table \ref{tab:par}. 
We calculate these parameters following the method in Appendix of
\citet{Nava97}.
The radial profiles of the ICM pressure is determined so that the ICM is
in hydrostatic equilibrium within the cluster potential of the DM and 
ICM itself. 
The resultant temperature profiles are similar to those of
initial cluster model in \citet{Rick01}.
The velocity distribution of the DM particles is assumed to be an isotropic
Maxwellian. The radial profiles of the DM velocity dispersion are 
calculated from the Jeans equation with spherical symmetry
so that the DM particles are in virial equilibrium in the cluster
potential.
The coordinate system is taken in such a way that the center of mass is 
at rest at the origin and that the $x-$axis is along with the collision
axis. The centers of the larger and smaller subclusters are initially
located in the sides of $x>0$ and $x<0$, respectively.
The initial distance between the subcluster's centers is 4.93 Mpc. 
The initial relative velocity is estimated as in \S 2 of
\citet{Taki99}. The resultant value is $8.98 \times 10^2$
km s$^{-1}$, which is approximately two thirds
of the infall velocity assuming that they
were at rest at infinite distance. 
\begin{table}
  \caption{Parameters for each subcluster}\label{tab:par}
  \begin{center}
    \begin{tabular}{llll}
      \hline \hline
       & $M_{DM}$ ($M_{\odot}$)& $r_{\rm v}$ (Mpc) & c \\
      \hline
      cluster 1 & $1.00 \times 10^{15}$   & 1.97   & 5.66 \\
      cluster 2 & $6.25 \times 10^{13}$   & 0.784  & 7.56 \\
      \hline
    \end{tabular}
  \end{center}
\end{table}

\section{The Simulation Results}
\begin{figure}
  \begin{center}
    \FigureFile(80mm,80mm){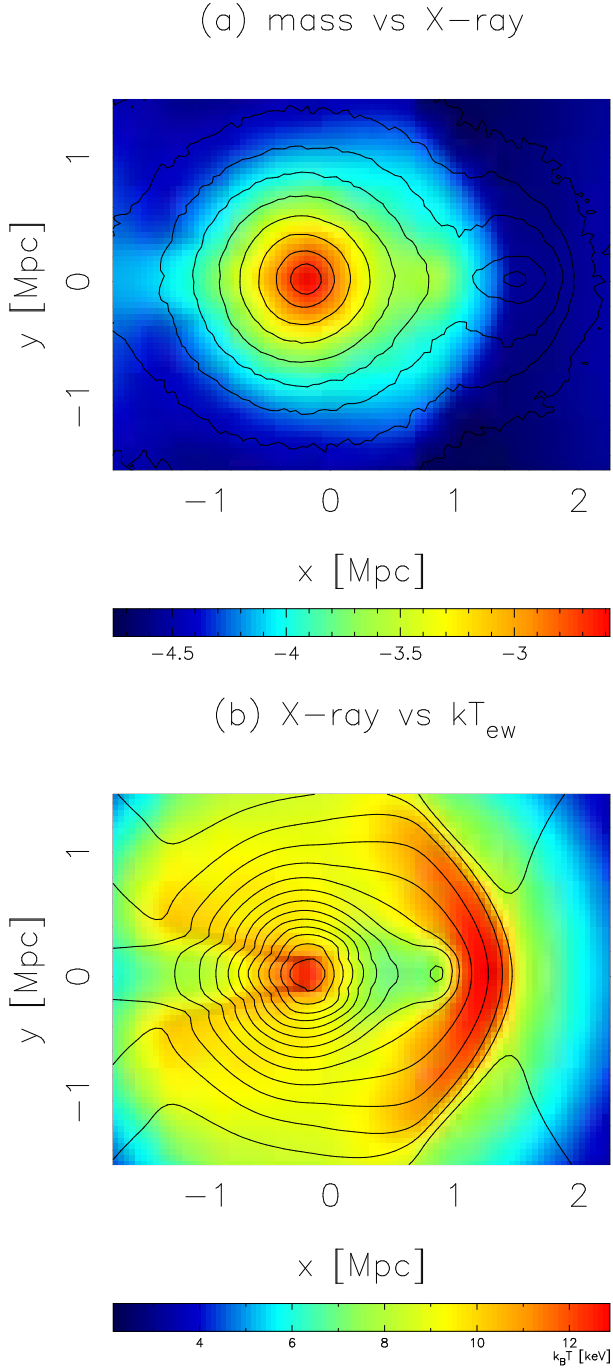}
  \end{center}
  \caption{(a)Projected total surface mass density (contours) overlaid
            with X-ray surface brightness distribution (colors) at a
            time of 0.67 Gyr after the core passage. A clear off-set of
            the mass density peak from the X-ray peak is seen.
           (b)X-ray surface brightness distribution (contours) overlaid
            with emissivity-weighted temperature distribution (colors)
            at the same epoch.}
   \label{fig:xray-dm}
\end{figure}

\begin{figure}
  \begin{center}
    \FigureFile(80mm,80mm){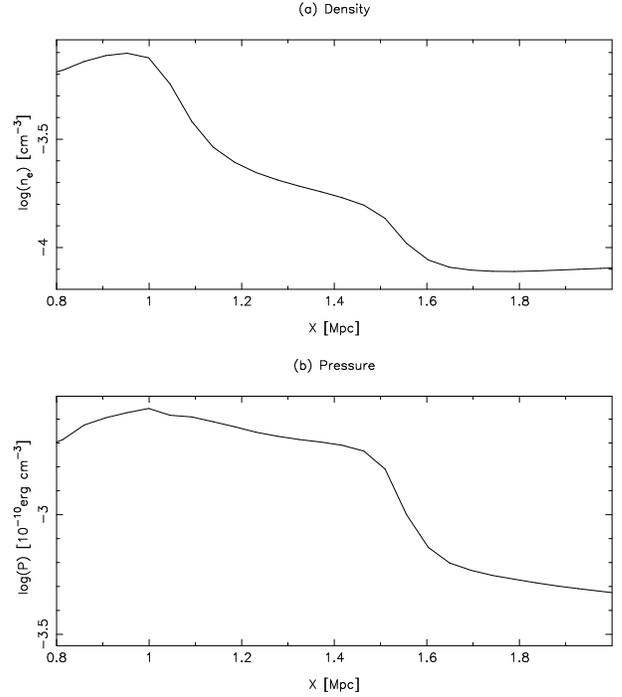}
  \end{center}
  \caption{Electron number density (a) and pressure (b) profiles along
           the collision axis ($y=z=0$) in front of the substructure at
           the same epoch as in figure \ref{fig:xray-dm}. A bow shock is
           located at $x \simeq 1.55$ Mpc, where jumps are clearly seen in
           both the density and pressure profiles. A contact discontinuity is 
           at $x \simeq 1.1$Mpc, where a jump is seen only in the density
           profiles and the pressure profile has no discontinuity there.}
   \label{fig:axis}
\end{figure}

Figure \ref{fig:xray-dm}(a) shows the X-ray surface brightness distribution
(colors) and projected total surface mass density (contours) 
at a time of 0.67 Gyr after the passage of the subcluster through the
core of the larger one. A clear off-set of the mass density peak from
the X-ray peak is seen
for the smaller subcluster remnant. The mass peak and X-ray peak are
located at $x \sim 1.5$ Mpc and $x \sim 1.0$ Mpc, respectively.
This is clearly because the ICM in the smaller subcluster is lagged 
by the ram pressure.
Figure \ref{fig:xray-dm}(b) shows the X-ray surface
brightness distribution (contours) overlaid with the emissivity-weighted
temperature distribution (colors) at the same epoch. A weak jump in the 
X-ray surface brightness distribution at $x \simeq 1.5$  
(near the smaller mass peak) is a
bow shock, and the emissivity-weighted temperature is higher in the 
brighter side, and vice versa. A more prominent jump in the X-ray 
brightness just in front of the smaller X-ray peak is a contact
discontinuity, and the emissivity-weighted temperature is lower in the
brighter side, and vice versa. Therefore, this will be recognized as a
cold front in actual X-ray observations.
Figure \ref{fig:axis}(a) and (b) show the ICM density and pressure
profiles along the collision axis ($y=z=0$) in front of the smaller X-ray
peak, respectively. 
The bow shock is located at $x \simeq 1.55$ Mpc, where jumps are clearly
seen in both the density and pressure profiles. The contact discontinuity is 
at $x \simeq 1.1$ Mpc, where a jump is seen only in the density profile
and the pressure profile does not have any discontinuity there. 
As for the overall ICM and DM structures of 1E 0657-56 around the 
west smaller X-ray and mass peak, our results agree qualitatively with 
the observations.

\section{Discussion on the Ram Pressure-Stripping Conditions}
Let us discuss the ram pressure-stripping conditions in merger of
two clusters with an NFW DM density profile.
We consider the merger of two clusters with
masses $M_1$ and $M_2$ ($M_1>M_2$), respectively. 
We concentrate on the physical status of the ICM in the core region
of the smaller subcluster. If the gravity on the
subcluster's ICM is weaker than the ram pressure force in
unit volume, the ICM will be stripped from the substructure potential. 
This means,
\begin{eqnarray}
  \frac{G m_2 \rho_2}{r_2^2} < 
          A (\pi r_2^2 \rho_1 v^2)(\frac{4}{3} \pi r_2^3)^{-1},
  \label{eq:condition}
\end{eqnarray}
where $G$ is the gravitational constant, and $\rho_1$ and $\rho_2$ are 
the central gas density of the subcluster
1 and 2, respectively. $r_2$ and $m_2$ are the scale radius of the DM
profile and the DM mass inside $r_2$ for the cluster 2, respectively.
Therefore, the relation between $m_2$ and $M_2$ is
\begin{eqnarray}
g(M_2) \equiv \frac{m_2}{M_2} = \frac{\ln2-1/2}{\ln(1+c)-c/(1+c)},
\end{eqnarray}
where, $c$ is a concentration parameter of an NFW profile,
and weakly depends on the halo mass. 
$A$ is a fudge factor of an order of unity. It is most
likely that $A \lesssim 1$ because all of the ram pressure force
is not effective in stripping the gas from the substructure. Some might be 
used on the excitation of small-scale eddies through Kelvin-Helmholtz
instability, and some on the adiabatic compression and shock heating of
the ICM, and so on.
The collision velocity $v$ has an order of
\begin{eqnarray}
  v^2 \simeq \frac{2G(M_1+M_2)}{R_1+R_2},
  \label{eq:velocity}
\end{eqnarray}
where $R_1$ and $R_2$ are the virial radii for the cluster 1 and 2, 
respectively.
It is convenient to introduce a new parameter $\alpha \equiv M_2/M_1$.
Then, the stripping condition of inequality (\ref{eq:condition}) becomes
\begin{eqnarray}
F(\alpha: M_1) \equiv \alpha^{2/3 - w} \frac{1+\alpha^{1/3}}{1+\alpha} - 
  \frac{3 A}{2 g(\alpha M_1) c(\alpha M_1)} < 0 \label{eq:condition2}.
\end{eqnarray}
Deriving inequality (\ref{eq:condition2}), we use the scaling relation that 
$R_2/R_1 = \alpha^{1/3}$ and $\rho_2/\rho_1 = \alpha^{-w}$.
In the $\Lambda$CDM universe ($\Omega_0=0.25$, $\lambda_0=0.75$), 
$w \simeq 1/4$ assuming that $\rho_1$ and 
$\rho_2$ behave like a characteristic density in an NFW
profile \citep{Nava97}.

\begin{figure}
  \begin{center}
    \FigureFile(80mm,80mm){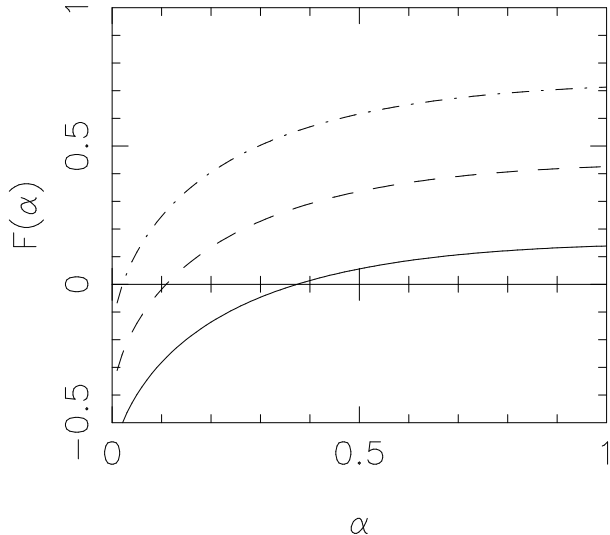}
  \end{center}
  \caption{Function $F(\alpha: M_1)$ defined by equation 
           (\ref{eq:condition2}) for $M_1=1.0 \times 10^{15} M_{\odot}$
           in the $\Lambda$CDM universe ($\Omega_0=0.25$,
            $\lambda_0=0.75$). The solid, dashed, and
           dot-dashed lines represent the cases of $A=0.6$,
           $0.4$, and $0.2$, respectively. $F<0$ means that the ram-pressure
           dominates the gravity in the substructure.}
   \label{fig:F(alpha)}
\end{figure}
Figure \ref{fig:F(alpha)} shows the function $F(\alpha: M_1)$ for
$M_1=1.0 \times 10^{15} M_{\odot}$ in the $\Lambda$CDM universe
($\Omega_0=0.25$, $\lambda_0=0.75$).
The solid, dashed, and dot-dashed lines represent the cases of $A=0.6$,
$0.4$, and $0.2$, respectively.
In any case, $\alpha$ is less than $\sim 0.1$ when $F(\alpha)<0$. 
This means that the ram pressure-stripping is more effective 
for smaller subclusters.
Although our estimation here is rather crude, it is interesting that this
criteria of $\alpha \sim 0.1$ is close to 
the mass ratio of our simulations where the clear off-set appears.
Obviously, such an off-set does not appear at all if the ram
pressure-stripping does not work effectively. 
The X-ray peak will correspond with the mass
peak because the ICM behaves like the DM. 
On the other hand, if the ram pressure overwhelms the gravity,
the ICM in the substructure will not be able to penetrate the larger
cluster's center and will be repelled. Furthermore, the larger cluster's
ICM is so hot that it cannot be bound by the substructure's
gravitational potential.
Therefore, we will see a mass peak associated with no X-ray peak. 
Clear off-sets in the simulation
and 1E 0657-56 suggest that the parameter $\alpha$ is close to 
the above-mentioned critical value where neither the ram pressure nor
the gravity absolutely dominates the other.

In our simulations, we do not take into account a dense cool core in
the central region of a cluster. However, this probably affects the ram
pressure stripping conditions. The existence of the cool core changes
$\rho_1$ and $\rho_2$ in inequality (\ref{eq:condition}). Thus, the
scaling relation $\rho_2/\rho_1=\alpha^{-w}$ could be modified and/or
have some dispersion. In case of 1E0657-56, the smaller subcluster
certainly has a cool core, but the larger one seems to have none. 
Therefore, $\rho_2/\rho_1$ can be larger than when neither subcluster has
a cooling core. This makes more difficult to strip the ICM.

Whereas the smaller X-ray peak in 1E0657-56 is triangle, that
in our simulation results is rather round. The absence of a cool core
might cause this difference. Another physical process that may play 
a crucial role is the magnetic field in the ICM.
The magnetic field along the boundary layer probably works 
to maintain the smaller subcluster's gas as a distinct structure 
after it is stripped off the DM potential through both 
the magnetic tension and suppression of heat conduction \citep{Asai04}.
Dynamical motion of the substructure itself possibly produces 
this kind of magnetic field configurations \citep{Vikh01}.
Furthermore, temperature gradients in the boundary layer 
might produce the magnetic field structure through Weibel-type
plasma instabilities \citep{Okab03}.
Three-dimensional high-resolution magnetohydrodynamic simulations 
will be useful in order to investigate detailed evolution. 

Please note that an off-set of an X-ray peak to a mass peak is 
not a structure in dynamical equilibrium but a transient one.
A characteristic timescale of the ram pressure-stripping  
is estimated to be $\sim R_2/v$, which becomes $0.42$ Gyr for the model
presented in \S 2. In the simulation, it takes $\sim 0.7$ Gyr
to strip the ICM from the substructure potential. Certainly
this is roughly equal to the above-mentioned value.

\section{Summary}
We investigate the X-ray and mass structures in the merging galaxy cluster
1E 0657-56. We first reproduce a clear off-set of an X-ray peak 
to a mass peak in $N$-body + hydrodynamical simulations of mergers of
two subclusters with an NFW and beta-model density profiles for DM and
ICM, respectively. As for the overall ICM and DM structures of 1E
0657-56 around the west smaller X-ray and mass peak, our simulation
results agree qualitatively with the observations. We discuss
the ram pressure-stripping conditions in the mergers of two clusters
with an NFW DM density profile using a simple analytical model. We find
that the ram pressure dominates the gravity of the substructure when the 
smaller cluster's mass is less than approximately one tenth of the
larger cluster's mass. The characteristic X-ray and DM structures found in
1E 0657-56 suggest that the mass ratio between the progenitors is close 
to the above-mentioned critical value and that neither the ram-pressure
force nor the gravity of the substructure overwhelms the other.

\bigskip
The author would like to thank T. Kitayama, M. Hattori and N. Okabe for 
valuable discussion, and H. Ishikawa for his contribution 
in the early stage of this work. 
The author is also grateful to an anonymous referee 
for his/her useful comments and suggestions
which improved the manuscript.
Numerical computations were carried out
on VPP5000 at the Center for Computational Astrophysics, CfCA, of the
National Astronomical Observatory of Japan.
M. T. was supported in part by a Grant-in-Aid from the
Ministry of Education, Science, Sports, and Culture of Japan (16740105). 


\end{document}